\documentstyle[prd,aps,preprint,graphicx]{revtex}

\begin{document}
\draft

%
%
\tightenlines

\preprint{Nisho-05/1}
\title{Color Ferromagnetism of Quark Matter
; a Possible Origin of Strong Magnetic Field in Magnetars}
\author{Aiichi Iwazaki}
\address{Department of Physics, Nishogakusha University, Ohi Kashiwa Chiba
  277-8585,\ Japan.} 
\date{Aug 15, 2005} \maketitle
\begin{abstract}
We show a possibility that strong ``magnetic field'' $\sim 10^{15}\,$G 
is produced by color ferromagnetic quark matter
in neutron stars. In the quark matter 
a color magnetic field is generated spontaneously owing to
Savvidy mechanism and a gluon condensate arises for 
the stabilization of the field.
Since the quark matter is electrically charged in the 
neutron stars, 
the rotation of the quarks around the color magnetic field
produces the strong 
``magnetic field''. 
\end{abstract}
\hspace*{0.3cm}
\hspace*{1cm}
\tightenlines

Observations of soft gamma-ray repeaters\cite{soft} and anomalous X-ray pulsars\cite{axp}
reveals an intriguing phenomenon associated with neutron star physics,
that is, the existence of extremely strong magnetic field $\geq 10^{15}\,$G.
The standard neutron stars\cite{neutron} possess also strong magnetic fields but
their strength is about $10^{12}\,$G, less by three order of magnitudes 
than the above one. Such neutron stars with the extremely strong magnetic 
field are called as magnetars. Both of soft gamma-ray repeaters and anormalous
X-ray pulsars have been considered to be magnetars.
The strength of the magnetic field
is estimated by spin down rates of pulsars
and there is a direct measurement\cite{direct} of the strength.
Furthermore,
there are some other reasons\cite{dun} supporting  
the strong magnetic field in the compact stars.             

Since the observed value $\sim 10^{15}\,$G of the magnetic field is the
one around surfaces of the stars,
its strength reaches $\sim 10^{18}\,$G in the cores, if the field
has a dipole-like configuration. 
Then, a naive question may arise: How is such a strong magnetic field produced ?
Conventionally, a dynamo mechanism\cite{dynamo} is believed
for the production
of the field.
We may, however, speculate from a simple energetical argument 
that quark matter causes the field. That is,
the typical neutron stars with surface magnetic field, $10^{12}\,$G,
would involve a nuclear or hadronic process generating the field 
whose energy scale 
is of the order of ten MeV. This energy scale comes from that of
the magnetic field $10^{15}\,$G present inside of the stars;
$\sqrt{10^{15}\,\mbox{G}}\simeq 8\,$MeV.
Similar consideration on the magnetar field leads to
a typical energy scale of QCD; $\sqrt{10^{18}\,\mbox{G}}\simeq 260\,$MeV. 
This indicates that the process generating the magnetar field
would be associated with quark matter, not hadronic matter.

In this paper we propose a possible mechanism for producing the strong magnetic field
in the magnetars. We assume that inside the stars there exists a phase
transition between dense hadronic matter and 
quark matter in a color ferromagnetic
phase ( CF phase ). Namely, the stars involve the quark matter
with a color magnetic field.
The CF phase has been discussed
in our previous papers\cite{cf1,cf2}. 
The point of the mechanism is that since a gas of quarks is both 
colored and electrically charged in the CF phase,
observable magnetic field is produced because of its rotation
around the color magnetic field.

We first give a brief review of the CF phase of the quark matter.
In the CF quark matter, the color magnetic field, $B$, is generated spontaneously
not by alignment of quark color spins, but by gluon' dynamics.
Namely, one loop effective potential of the color magnetic field 
has non trivial minimum at $B \neq 0$. This comes 
from the quantum effects of
gluons.
Thus, there is a possible
CF ground state of gluons.
This is the original 
analysis by Savvidy\cite{savvidy}. Since the loop approximation is valid
at large baryon chemical potentials, 
namely, at small gauge coupling constant, 
the CF phase may arise
in the dense quark matter.
Usually, di-quark condensation
is taken into account and 
only color superconducting phase ( CS phase ) is discussed\cite{color}
in the dense quark matter.
But, in order to
find possible phases of the dense quark matter,
the CF state should be 
taken into account.
Obviously, these two phases are incompatible.
Hence we have compared\cite{cf2} free energies of quarks in each phase
in order to find
which phase is favored. We have found that
the CF phase is more favored than  
the CS phase at lower baryon chemical potentials.
On the other hand, the CS phase is
more favored than the CF phase at higher baryon chemical potentials.
This result holds only at the extremely large chemical potential
so as for the loop approximation to be valid. 
In the paper we assume that the result may hold even at
smaller chemical potentials at which the phase transition
occurs from hadronic matter to the quark matter.

Roughly speaking, hadronic phase is realized due to the condensation of color 
magnetic monopoles. At large gauge coupling constant, $g$, interactions between
the monopoles, $\sim 1/g^2$, 
are much small so that almost free gas of the monopoles may condense. Thus,
the quark confinement arises due to the realization of
color magnetic superconducting state\cite{confinement,abelian}. With decreasing the gauge coupling
constant, the interactions between the monopoles increase.
Consequently, the dipoles of the
monopoles are formed and the condensation melts down. Probably, their dipole moments are
aligned so that the color magnetic field is produced.
This is a naive physical picture of the CF phase. 

It seems apparently that the quarks do not play any roles 
for the realization of the CF phase in the above argument
except for yielding small gauge coupling constant.
We have shown\cite{cf1} that the quarks play an important role 
for the stabilization of the color magnetic field.
Namely,  
unstable modes of gluons, which are present\cite{unstable} under the 
color magnetic field,  
have been shown to be stabilized with their condensation,
just as scalar fields in Higgs potentials.
The condensation leads to 
a fractional quantum Hall state of the gluons\cite{qh,cf1} with 
a color charge density. This color charge density
of the gluons is supplied by quarks. That is,
the quarks give color charges for the gluon sector necessary
for the formation of the quantum Hall state.  
This is a role of the quarks for the realization of the CF phase.
Consequently, the gas of the quarks in the quark matter
is charged in color, which results in the production of
the observable magnetic field.
   
Now, we explain the production mechanism of the strong magnetic fields
observed in magnetars. For the purpose, we
show in more detail 
how the unstable gluons under the color magnetic 
field are stabilized. 
We consider SU(2) gauge theory with massless quarks
of the two flavors for simplicity.
 
Taking the direction of $B$ in color space being in
$\lambda_3=\sigma_3/2$ ( $\sigma_i$ denotes Pauli metrices ),
we decompose the gluon's
Lagrangian
with the use of the variables, ``U(1) gauge field" 
$A_{\mu}=A_{\mu}^3,\,\,\mbox{and} \,\, \mbox{``charged vector field"}\,
\Phi_{\mu}=(A_{\mu}^1+iA_{\mu}^2)/\sqrt{2}$ 
where indices $1\sim 3$ denote color components,

\begin{eqnarray}
\label{L}
L=-\frac{1}{4}\vec{F}_{\mu
  \nu}^2&=&-\frac{1}{4}(\partial_{\mu}A_{\nu}-\partial_{\nu}A_{\mu})^2-
\frac{1}{2}|D_{\mu}\Phi_{\nu}-D_{\nu}\Phi_{\mu}|^2 \nonumber \\
& & \ \ +ig(\partial^{\mu}A^{\nu}-\partial^{\nu}A^{\mu})\Phi_{\mu}^{\dagger}
\Phi_{\nu}+\frac{g^2}{4}(\Phi_{\mu}\Phi_{\nu}^{\dagger}-
\Phi_{\nu}\Phi_{\mu}^{\dagger})^2
\end{eqnarray}
with $D_{\mu}=\partial_{\mu}+igA_{\mu}$.
We have omitted a gauge term $D_{\mu}\Phi^{\mu}=0$. 
Using the Lagrangian we can derive that the energy $E$ of
the charged vector field $\Phi_{\mu}\propto e^{iEt-ikx_3}$ 
in the magnetic field, $A_{\mu}=A_{\mu}^B$, is given by
$E^2=k^2+2gB(n+1/2)\pm 2gB$  
with a gauge choice, $A_j^B=(0,x_1 B,0)$ and $(\partial_{\mu}+igA_{\mu}^B)\Phi^{\mu}=0$, 
where we have taken the spatial direction of $\vec{B}$ being along $x_3$ axis.
$\pm 2gB$ ( the integer $n\geq 0$ ) denotes the contribution from spin components
of $\Phi_{\mu}$ ( Landau levels )
and $k$ denotes
momentum in the direction parallel to the magnetic field.

We find that unstable modes of gluons are given by 
$\Phi=(\Phi_1-i\Phi_2)\sqrt{1/2}$ occupying the Lowest Landau level, $n=0$
and that their spectra are given by $E^2=k^2-gB$, which are
negative for $k^2<gB$. Thus, their amplitudes increase
rapidly in time; $\Phi\sim e^{|E|t}$. In other words, they are excited
spontaneously and form eventually
a stable ground state owing to the self interactions, $g^2\Phi^4$.
Nielsen, et al.\cite{flux} tried to find a stable configuration of
the mode $\Phi$ with $k=0$, which minimizes a classical
potential energy of $\Phi$, $V(\Phi)=-2gB\Phi^2+g^2\Phi^4/2$;
the form of $V(\Phi)$ can be derived by taking only the 
unstable modes, $\Phi$ in the Lagrangian.
The configuration they obtained is ``flux lattice'' of $\Phi$,
and resultant magnetic field is given by $F_{12}=B-g^2|\Phi^2|$.
Namely, they found that $|\Phi(x_1,x_2)|$ is periodic in two spatial coordinates, $x_1$ and
$x_2$. Their periodicity is approximately given by 
the magnetic length, $l_B=1/\sqrt{gB}$. Since fractional quantum 
Hall states\cite{qh} had not been known before 1982, it was a reasonable 
solution they could obtain. 
At present we know 
that even bosons such as the unstable gluons,
$\Phi$ with $k=0$, 
may form stable fractional quantum Hall states\cite{boson} when they 
occupy the Lowest Laundau level, interacting
repulsively with each others.
The states are characterized by the following filling factor, $\nu$,

\begin{equation}
\nu=\frac{2\pi \rho_2}{gB}=\frac{1}{2\times \mbox{positive integer}}
\end{equation}  
with two dimensional color charge density, $\rho_2$ of the gluons.
The filling factor is defined by the ratio of
the number ( charge ) density of gluons to the degeneracy per unit
area of each Landau level; $\nu=\mbox{``density''}/(gB/2\pi)$.
We note that even integers in the denominator appear because of the
bosons; odd integers appear in the case of Fermions ( electrons ).
Here we consider only the case of $\nu=1/2$.
The detail how the unstable gluons form the quantum Hall state
should be addressed to our previous papers\cite{cf1,cf2}.
( We have demonstrated the formation of the gluon's quantum Hall states
by using Chern-Simons gauge theory, which have been
originally applied\cite{cs} by us for understanding field theoretically Laughlin 
states of electrons. )  
We have shown\cite{cf1} that this quantum Hall state with $\nu=1/2$ is energetically
more favored than the ``flux lattice'', which may be regarded as 
a Wigner crystal\cite{qh} of
the gluons in the modern point of view.   
In this way the unstable gluons under the color magnetic field
are stabilized by
forming the quantum Hall states.

Then, we must ask how large the coherent length of the color
magnetic field in $x_3$ direction is. Namely, we must ask
how large the width of two dimensional ``quantum well''\cite{qh} is.
The quantum Hall states are formed in the well.  
Here we give only a plausible argument about the width,
although more rigorous treatment is necessary.
Since the unstable modes of
gluons disturb the coherence of the magnetic field, their excitations
with momenta $k<\sqrt{gB}$ make the coherent length diminish 
to be $l_B=1/\sqrt{gB}$. ( The interpretation is consistent with the
fact\cite{cf1} that the condensation of the gluons gives a mass $\sim \sqrt{gB}$
to the magnetic field. )
This implies that the quantum Hall state is realized effectively in a
quantum well with its width, $1/\sqrt{gB}$. This well is
extending infinitely in $x_1$ and $x_2$ directions in quark matter.
There exist many these wells perpendicular to $\vec{B}$
in the quark matter.
The direction of each
color magnetic field in the wells is aligned to a direction of
e.g. $x_3$. 
Consequently, the
color magnetic field  
may exist globally in the quark matter, although the real coherence
of the field is restricted within a well.
Since the two dimensional color charge density $\rho_2$ 
necessary for the formation of the
quantum Hall state is localized in the well, 
three dimensional color charge
density, $\rho_3$, is given by

\begin{equation}
 \rho_3\simeq \rho_2/l_B=\rho_2\sqrt{gB}=\frac{(gB)^{3/2}}{4\pi}.
\end{equation}

This charge density is carried by the gluons. 
Since the quark matter in compact stars is
color neutral,
the color charge of the gluons is compensated by quarks.
Thus, the gas of the quarks becomes to possess a color charge density of
$-\rho_3$. 
Since the quark matter is not electrically neutral,
its rotation around the color magnetic field can produce the 
observable magnetic field.
This is the physical origin in our model for the generation
of the observable strong magnetic field.

Now, we calculate the strength of the observable magnetic field produced spontaneously in 
the quark matter, that is, spontaneous magnetization of the matter. 
The point is that the number of negative color charged quarks
is different from the number of positive color charged quarks.
The difference induces 
an electric current around the color magnetic field.
We consider only u and d quarks.
Number densities  and energy densities of the quarks are given by

\begin{equation}
n_{u,d}^{\pm}(gB)=\frac{gB\mu_{u,d}^{\pm}}{4\pi^2} \quad \mbox{and} \quad \epsilon_{u,d}^{\pm}(gB)=\frac{gB(\mu_{u,d}^{\pm})^2}{8\pi^2}
\end{equation}  
where $\mu_{u,d}^{\pm}$ are chemical potentials of each flavor of
quarks with $\pm$ color charges associated with the generator
$\lambda_3$ of SU(2) gauge group. 
We assume for simplicity  that all of quarks occupy only 
the lowest Landau level, that is, Fermi energy $\mu$
is less than $\sqrt{gB}$. In order to obtain free energies,
we notice three conditions\cite{cf2,condition} which must be satisfied in the
neutron stars; 
the conditions of color and electric neutralities, and of
beta-equilibrium ( u $\leftrightarrow$ d$+$e$^{-}$ ).  

\begin{eqnarray}
\rho_3&=&\frac{(gB)^{3/2}}{4\pi}=(n_u^{-}+n_d^{-}-n_u^{+}-n_d^{+})/2,\nonumber \\
0&=&2(n_u^{+}+n_u^{-})/3- (n_d^{+}+n_d^{-})/3-n_e,\nonumber \\
\mu_d^{\pm}&=&\mu_u^{\pm}+\mu_e,
\end{eqnarray}
where $n_e=\mu_e^4/(4\pi^2)$ is the number density of electrons
with the chemical potential denoted by $\mu_e$.

In order to obtain the magnetization, $M$, we need to estimate
the free energy,
$G(H)$; $M=-\partial_{H}G(H)$ for $H\to 0$, where
$H$ is an external magnetic field coupled with electric charges.
Assuming that the direction of $\vec{H}$ points to the one of the color magnetic
field, $\vec{B}$, we obtain

\begin{eqnarray}
G(H)&=&\epsilon_u^{+}(gB+2eH/3)+\epsilon_u^{-}(gB-2eH/3)+\epsilon_d^{+}(gB-eH/3)+\epsilon_d^{-}(gB+eH/3)
\nonumber \\
&-& \Bigl( \mu_u^{+}n_u^{+}(gB+2eH/3)+\mu_u^{-}n_u^{-}(gB-2eH/3)+\mu_d^{+}n_d^{+}(gB-eH/3)
\nonumber \\
&+&\mu_d^{-}n_d^{-}(gB+eH/3) \Bigr) \nonumber \\
&\simeq&-\frac{gB((\mu_u^{+})^2+(\mu_u^{-})^2+(\mu_d^{+})^2+(\mu_d^{-})^2)}{8\pi^2}-
\frac{eH(2(\mu_u^{+})^2-2(\mu_u^{-})^2-(\mu_d^{+})^2+(\mu_d^{-})^2)}{24\pi^2},
\end{eqnarray} 
where we have neglected the contribution of electrons because they do
not produce any magnetization in the limit of $H \to 0$. Thus, the
magnetization is 

\begin{eqnarray}
M&=&e\frac{2(\mu_u^{+}+\mu_u^{-})(\mu_u^{+}-\mu_u^{-})-(\mu_d^{+}+\mu_d^{-})(\mu_d^{+}-\mu_d^{-})}{24\pi^2}\nonumber 
\\
 &=&e\frac{(\mu_u^{+}+\mu_u^{-}-2\mu_e)(\mu_u^{+}-\mu_u^{-})}{24\pi^2}
\end{eqnarray}
where we have used the beta-equilibrium condition.
Since the difference of $\mu_u^{+}-\mu_u^{-}$ can be represented by
the color charge density of the quarks, $-\rho_3$, we find that

\begin{eqnarray}
M&=&-e\frac{\sqrt{gB}\,(\mu_u^{+}+\mu_u^{-}-2\mu_e)}{24\pi}\nonumber \\
&\simeq & -\frac{e\pi n_B}{4\sqrt{gB}}\sim 10^{17}\,\mbox{G}\,
\frac{400\,\rm{MeV}}{\sqrt{gB}}\frac{n_B}{1/\rm{fm}^3}
\end{eqnarray} 
where $n_B$ denotes the baryon number density of the quarks;
$n_B=(n_u^{+}+n_u^{-}+n_d^{+}+n_d^{-})/3=gB(\mu_u^{+}+\mu_u^{-}+\mu_e)/6\pi^2$ .
We have neglected the negligible contribution of electrons to $M$. 
It turns out that the magnetic field ( magnetization ) obtained
is sufficiently large to explain the strong magnetic field 
of the magnetars.

We have not yet determined the value of $gB$.
The value should be obtained by using the phenomena associated with the color
ferromagnetism of the quark matter; there are still not such phenomena.
But we may expect that
it takes about typical QCD scale such as several hundred MeV; it does
neither several ten MeV or several GeV. 
Therefore, 
if the radius of the quark matter in compact stars is about $2$ km,
the strength of the surface magnetic field reaches at 
$10^{15}\,$G. 
In this way, we explain that the extremely strong
magnetic field of the magnetars is caused by
the quark matter in the CF phase.
Thus, the magnetars may involve the quark matter
inside their cores.
( We have presented\cite{cf2} a structure of a neutron star
involving both the
CF quark matter and nuclear matter by using
appropriate equations of state in both matters. )   

We have argued that a kind of quantum wells
in which quantum Hall states of gluons are present are
formed effectively. The argument is based on the fact that
the unstable modes, $\Phi(k)$ with momenta, $k<\sqrt{gB}$, render 
the coherent length of the color magnetic field, $F_{12}=B-g^2|\Phi(k)|^2$,  
diminish. As a result, a quantum well may be formed
in which the color magnetic field with the diminished coherent
length $\simeq l_B$ is present. This formation of the well
is essential to obtain the observable strong magnetic field,
since the color charge density $\rho_3\simeq \rho_2/l_B$ of the quark
matter is used to derive the magnetization.
The argument is plausible, but never rigorous. 
We wish to make the point more clearly in the future.

We have estimated the spontaneous magnetization of 
the CF quark matter 
in SU(2) gauge theory. We have neglected the effects of the strange quarks and 
of higher Landau levels occupied by quarks.
It is straightforward to include their effects in SU(3) gauge
theory. 
It is expected that even if we include all of the effects,
the
strength of the magnetic
field will be of the same order of magnitude as the ones 
estimated in this paper. We will report it in near future.

\vspace*{2em}
We would like to express thanks
to Prof. O. Morimatsu, Dr. T. Nishikawa
and Dr. M. Ohtani for useful discussion.
This work was supported by Grants-in-Aid of the Japanese Ministry
of Education, Science, Sports, Culture and Technology (No. 13135218).


\end{document}